\documentclass[12pt]{article}
\usepackage{verbatim}
\usepackage{amsfonts}
\usepackage{graphics}
\usepackage{amsmath}
\usepackage{times}
\usepackage{appendix}
\usepackage{color}
\usepackage{soul}
\usepackage{mathtools}
\usepackage{enumerate}
\usepackage{fancyhdr,latexsym,amsmath,amsfonts,amssymb,amsbsy,amsthm,url}
\usepackage{amsthm,amssymb,mathrsfs,setspace,pstcol,pstricks}
\usepackage[margin=0.5in,footskip=0.25in]{geometry}
\usepackage{graphics,graphicx,epsfig}
\usepackage{breqn}
\usepackage{caption,subcaption,float,subfloat}
\usepackage{pdflscape}
\usepackage{hyperref}
\usepackage{enumitem}
\usepackage[ruled,vlined]{algorithm2e}

\usepackage[english]{babel}
\usepackage[dvipsnames]{xcolor}

\usepackage{tfrupee} 
 
\makeatletter

\renewcommand{\section}{
	\@startsection
	{section}
	{1}
	{0pt}
	{1.1\baselineskip}
	{0.2\baselineskip}
	{\sc \centering}
}

\renewcommand{\subsection}{
	\@startsection
	{subsection}
	{1}
	{0pt}
	{1.1\baselineskip}
	{0.2\baselineskip}
	{\sc \centering}
}

\renewcommand{\subsubsection}{
	\@startsection
	{subsubsection}
	{1}
	{0pt}
	{1.1\baselineskip}
	{0.2\baselineskip}
	{\sc \centering}
}

\makeatother

\usepackage[flushleft]{threeparttable}
\usepackage{rotating,booktabs,multirow}
\usepackage{colortbl}
\usepackage{makecell,cellspace,caption}

\begin{document}
	
\title{\large\sc Green portfolio optimization: A scenario analysis and stress testing based novel approach for sustainable investing in the paradigm Indian markets}
\normalsize
\author{\sc{Shashwat Mishra} \thanks{Department of Mathematics, Indian Institute of Technology Guwahati, Guwahati-781039, India, e-mail: shashwat.mishra@iitg.ac.in}
\and 
\sc{Rishabh Raj} \thanks{Department of Mathematics, Indian Institute of Technology Guwahati, Guwahati-781039, India, e-mail: rishabh.raj@iitg.ac.in}
\and 
\sc{Siddhartha P. Chakrabarty} \thanks{Department of Mathematics, Indian Institute of Technology Guwahati, Guwahati-781039, India, e-mail: pratim@iitg.ac.in}
}

\date{}
\maketitle
\begin{abstract}

In this article, we present a novel approach for the construction of an environment-friendly green portfolio using the ESG ratings, and application of the modern portfolio theory to present what we call as the ``green efficient frontier'' (wherein the environmental score is included as a third dimension to the traditional mean-variance framework). Based on the prevailing action levels and policies, as well as additional market information, scenario analyses and stress testing are conducted to anticipate the future performance of the green portfolio in varying circumstances. The performance of the green portfolio is evaluated against the market returns in order to highlight the importance of sustainable investing and recognizing climate risk as a significant risk factor in financial analysis.


\end{abstract}

\section{Introduction}
\label{Sec_Introduction}

In the context of societal and existential crisis (resulting from climate change) in general and sustainable finance in particular, one can categorize the genesis of the financial risks (as a direct consequence of climate change)
into two components, namely physical risks and transition risks \cite{Hellmich2021}. The drivers of physical risk include extreme weather events and progressive change in the climate, while the transition risk can be attributed to three broad factors, namely, changes in regulatory mechanisms, choice and pace of technology adoption, and public sentiment. In \cite{Hellmich2021}, the authors identify taxation and Emission Trading Schemes (ETS) as two possible solutions to mitigate the financial risk arising out of climate change, and observe that the latter (ETS) is the preferred approach for the same. The ETS can be viewed as an action that is adopted and is incumbent upon the firms. On the other hand, from the perspective of investors and stakeholders of the financial system, a common adopted risk mitigation approach includes the Environmental Social and Governance (ESG) score \cite{Roncalli2020}.

In the context of sustainable investment, several approaches towards the development of sustainable portfolio strategies have appeared in literature, in terms of the classical Markowitz as well as the Capital Asset Pricing Model (CAPM) framework. For the former, ``decarbonization'' is a terminology used in the context of design of sustainable portfolios \cite{Koch2013}. One approach introduced in \cite{Andersson2016} is based on the problem of minimization of the tracking error, as given by the standard deviation of the differences in the returns of a green (or decarbonized) portfolio and a benchmark index (or portfolio). Further, one can also introduce carbon intensity (both revenue based and profit based) in the process of management of sustainable portfolios \cite{Fang2019}. The traditional concept of the Sharpe ratio frontier has been extended to incorporate ESG score of the portfolio components in \cite{Pedersen2021}. In a recent work, a CAPM based formula for assets, in presented through the inclusion of a Brown Minus Green (BMG) factor \cite{Roncalli2021}.

According to \cite{ClimateTracker}, the Nationally Determined Contribution (NDC) for India, has set a unconditional NDC target in COP26, for 2030, to reduce its emissions intensity by $45\%$ of the 2005 levels. 
Further, as a part of 2030 conditional NDC target, the goal is to increase the non fossil-fuel based electric power capacity to be at $50\%$ of the cumulative electric power capacity, with the eventual goal of net-zero target by 2070. Consequently, for investors in the Indian market, there is now an emergent transition risk, which now has to be accounted for in the financial investment strategies. Further the potential impact of both physical and transition risks, in the paradigm of Indian markets have to be subjected to both scenario analysis and stress tests
\cite{Roncalli2020,De2004,Breuer2013,Breuer2020}, which is the main focus of our undertaking the study reported in this article.

For the purpose of estimating the environmental impact in case of Indian financial system, this paper considers the environment score of various companies as published by CRISIL \cite{CRISIL_Ratings}. Using the data, we propose a novel method for portfolio optimization by visualizing not only the risk versus return, but also the suitably calculated environment score for the portfolio, resulting in formation of a novel green portfolio. Further, in order to conduct stress testing, we undertake the construction of a linear model by extending the capital asset pricing model and including factors related to physical and transition risk. Using this model, we proceed with the stress testing for the green portfolio by considering various scenarios which are likely to occur in the future. These scenarios take into account the targets for carbon emissions set by India, and also the possibility of a natural disaster which causes high economic damages, resulting in adverse impact on the market.

\section{Constructing a green portfolio}

In this section, we focus on the problem of constructing a green portfolio of equities. We first begin with listing out the data sources that were used in this work. Since this work is focused on Indian equity markets, we therefore, restrict ourselves to stocks that are listed in the National Stock Exchange (NSE) of India. Firstly, the stock price data for every company rated by CRISIL \cite{CRISIL_Ratings} and included in the ESG Ratings was obtained from Google Finance directly into Google Sheets through the GOOGLEFINANCE function \cite{GOOGLE_Finance}. The TICKER symbol data for all the listed companies was downloaded from the NSE's publicly available online data \cite{NSE_India}. This data was used to retrieve the TICKER symbols for companies in ``CRISIL's ESG Scores 2022'' list of companies by matching company names using Excel formulae. Hence, company identification through the TICKER symbol allowed for the extraction of company data through the GOOGLEFINANCE functionality.

As a part of the asset picking exercise, a total of 25 companies are chosen, with the goal of keeping the portfolio diversified but without having to include too many assets. We collate and use the returns data for the period of 1999-2023, and accordingly include only those companies who data was available for the entire duration. The list of companies was further narrowed down by filtering out the companies with environmental scores less than or equal 35 (out of 100), which left us with a total of 137 companies. We chose the companies (for inclusion in the portfolio) in such a way that almost $60\%$ of the total companies in the portfolio were large-cap, $25\%$ were mid-cap, and $15\%$ were small-cap companies. Consequently, our portfolio comprised of 15 large-cap, 6 mid-cap and 4 small-cap companies from the list. A selection process was then adopted, using multiple metrics like 24-Year CAGR, 3-Year CAGR and Excess Returns over Market to narrow down a list of 25 well-performing companies from 18 different sectors, with an impressive average environment score of 57.28.

\subsection{Description of the portfolio}

The portfolio (of 25 stocks) under consideration, is defined in terms of their weights as,
\[w_{i}=\frac{x_{i}S_{i}(0)}{V(0)},~\text{for}~i=1:25,\]
that is, the weight of the $i$-th asset is denoted by $w_{i}$ ($x_{i}$, $S_{i}(0)$ and $V(0)$ are the number of units of $i$-th asset, current price of the $i$-th asset and the current value of the portfolio, respectively). Further, let $\mu_{i}$ be the expected return on the $i$-th asset and $c_{i,j}$ be the covariance between the returns of the $i$-th and $j$-th asset. Finally, we introduce the notation for the environmental score of the $i$-th asset as $es_{i}$. At this juncture, we enumerate the relevant vectors and matrix as follows:
\begin{enumerate}[label=(\Alph*)]
\item Weight vector: $\displaystyle{\mathbf{w}=\begin{bmatrix} w_{1} & w_{2} & \dots & w_{25} \end{bmatrix}}$.
\item Unit vector: $\displaystyle{\mathbf{u}=\begin{bmatrix} 1 & 1 & \dots & 1 \end{bmatrix}}$.
\item Expected return vector: $\displaystyle{\mathbf{m}=\begin{bmatrix} \mu_{1} & \mu_{2} & \dots & \mu_{25}\end{bmatrix}}$.
\item Environmental score vector: $\displaystyle{\mathbf{es}=\begin{bmatrix} es_{1} & es_{2} & \dots & es_{25}
\end{bmatrix}}$.
\item Covariance matrix: $\displaystyle{\mathbf{C}=\begin{bmatrix}
c_{1,1} & c_{1,2} & \dots & c_{1,25}\\ c_{2,1} & c_{2,2} & \dots & c_{2,25}\\
\vdots & \vdots & \ddots & \vdots \\ c_{25,1} & c_{25,2} & \dots & c_{25,25}
\end{bmatrix}}$.
\end{enumerate}
This leads us to the following constraint and parameters for the portfolio $V$:
\begin{enumerate}[label=(\Alph*)]
\item Constraint on weights: $\displaystyle{\mathbf{u}\mathbf{w}^{\top}=1}$.
\item Portfolio expected return: $\displaystyle{\mu_{V}=\mathbf{m}\mathbf{w}^{\top}}$.
\item Portfolio risk: $\displaystyle{\sigma^{2}_{V}=\mathbf{w}C\mathbf{w}^{\top}}$.
\item Environmental score for the portfolio: $\displaystyle{es_{V}=\mathbf{es}\mathbf{w}^{\top}}$.
\end{enumerate}

\subsection{Generation and visualization of attainable portfolios}

\begin{figure}[h]
\begin{center}
\includegraphics[scale=0.75]{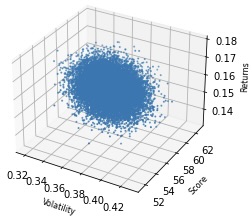}
\caption{Attainable portfolios}
\label{fig 5.1}
\end{center}
\end{figure}

\begin{figure}[h]
\begin{center}
\includegraphics[scale=0.75]{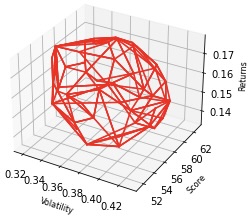}
\caption{Green efficient frontier}
\label{fig 5.2}
\end{center}
\end{figure}

Firstly, random portfolio weights with $\mathcal{U}(0,1)$ distribution are generated and then normalized so that their sum is $1$. It may be noted that negative weights are not considered in this exercise, thereby ensuring that there is no short-selling. Each set of 25 weights is used to define a single portfolio. For each portfolio, the expected return, volatility (standard deviation) and the environment score are calculated. These three points thus obtained characterizes a portfolio. For the purpose of our work, we generated 20,000 such portfolios which will be used as our data points. These data points were used to create a 3D scatter plot as depicted in Figure \ref{fig 5.1}, showing the characteristics of all the portfolios. We can observe that the figure approaches a closed shape which encloses all the possible portfolios that can be constructed (that is, all admissible portfolios). The surface which enclosed the 3D scatter plot represents the efficient frontier and we call this as the ``green efficient frontier''. In order to plot it (the green efficient frontier) numerically, portfolios depicted in Figure \ref{fig 5.1} are considered and a convex hull is formed. The outermost portfolio points are taken from the figure and used to create the corresponding 3D frontier as a bounding box, as shown in Figure \ref{fig 5.2}. 

\begin{figure}[h]
\begin{center}
\includegraphics[scale=0.75]{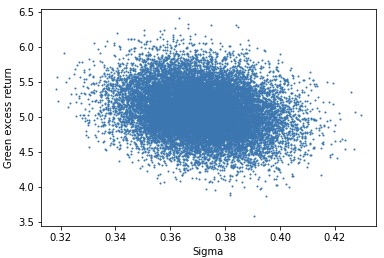}
\caption{Green excess return vs $\sigma$}
\label{fig 5.3}
\end{center}
\end{figure}

While selecting the optimal portfolio, there is a threefold objective to be meet as enumerated below:
\begin{enumerate}[label=(\Alph*)]
\item Maximize the difference between expected portfolio return and risk-free return.
\item Minimize volatility of the overall portfolio.
\item Maximize the environment score of the portfolio.
\end{enumerate}
Accordingly, in order to achieve these objectives, the maximization of a custom metric is considered, which is called the ``green excess return''. This is defined as:
\[(\mu_{V}-r_{f})es_{V}.\] 
Here $r_{f}$ stands for the risk-free rate, whose value is chosen to be $r_{f}=6.950\%$, which is the yield on a one-year bond issued by the Government of India. Plotting the green excess return against $\sigma$ for each of the portfolios, Figure \ref{fig 5.3} is obtained.

\begin{figure}[h]
\begin{center}
\includegraphics[scale=0.75]{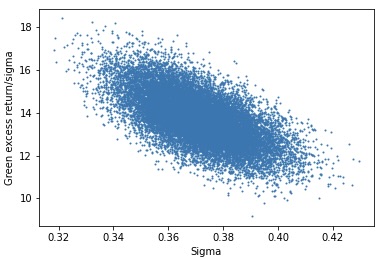}
\caption{Green excess return scaled by $\sigma$ vs $\sigma$}
\label{fig 5.4}
\end{center}
\end{figure}

Now, the portfolio with the highest ratio of green excess return per unit of sigma is considered. This corresponds to maximizing the following: 
\[\left[\frac{\mu_V-r_f}{\sigma}\right]\times es_V.\] 
The term within brackets is the Sharpe ratio, which is a financial measure that helps investors evaluate the risk-adjusted return of an investment. Specifically, it measures the excess return earned by an investment (its return above the risk-free rate) per unit of risk taken (its standard deviation). The Sharpe ratio is scaled by multiplying it with the environment score of the portfolio to give the annual green excess return. For this, another graph is plotted where this metric is shown for each of these portfolios compared with the $\sigma$. This can be seen in Figure \ref{fig 5.4}. Here, it can be observed that the highest ratio for a portfolio is above $18$, and the corresponding $\sigma$ is slightly higher than the minimum $\sigma$. This portfolio is selected as the green portfolio.

\subsection{The dynamics for the green portfolio}

\begin{figure}[h]
\begin{center}
\includegraphics[scale=0.75]{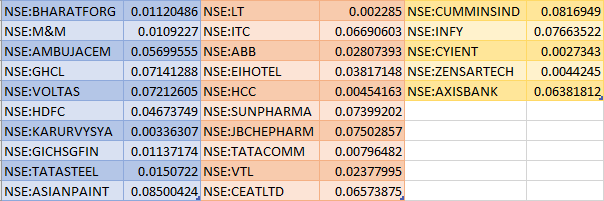}
\caption{Optimal weights}
\label{fig 5.5}
\end{center}
\end{figure}

\begin{figure}[h]
\begin{center}
\includegraphics[scale=0.75]{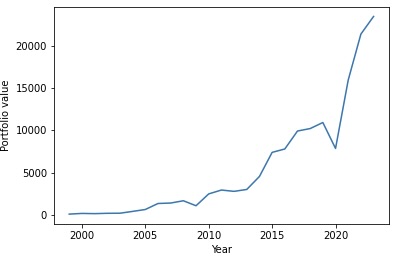}
\caption{Evolution of the optimal green portfolio}
\label{fig 5.6}
\end{center}
\end{figure}

\begin{figure}[h]
\begin{center}
\includegraphics[scale=0.75]{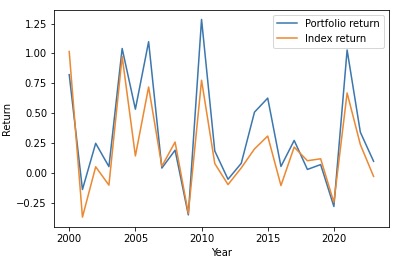}
\caption{Comparison of green portfolio returns with BSE-100}
\label{fig 5.7}
\end{center}
\end{figure}

The optimal weights obtained are shown in Figure \ref{fig 5.5}. Assuming that an amount of \rupee~$100$ was invested in this portfolio in 1999, the evolution of the portfolio value is presented in Figure \ref{fig 5.6}.
Finally, the green portfolio's historical returns are compared with the Bombay Stock Exchange (BSE) 100 market index returns (BSE was chosen as a representative of the market, different from NSE, from where the stocks of the portfolio were chosen). For this, the returns over time for both portfolios are plotted on the same figure, namely, Figure \ref{fig 5.7}.

\section{Green portfolio returns model}

\begin{figure}[h]
\begin{center}
\includegraphics[scale=0.8]{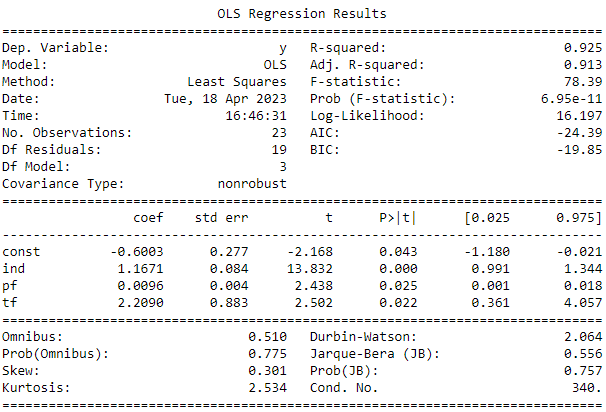}
\caption{Result of linear regression on the green portfolio returns model}
\label{fig 6.1}
\end{center}
\end{figure}

In this section, we extend the the concept of capital asset pricing model (CAPM) in the paradigm of green portfolios. Accordingly, we assume that there exists a linear and additive relation between the excess market return, physical risk factor, transition risk factor, and the portfolio return. Thus, we model our green portfolio returns as,
\[r_{V}=r_{f}+\alpha +\beta_{1}\left(r_{m}-r_{f}\right)+\beta_{2}PF+\beta_{3}TF.\]
Here,
\begin{enumerate}[label=(\Alph*)]
\item Portfolio return: $r_{V}$.
\item Risk-free rate of return: $r_{f}$ (as already defined earlier).
\item Market return: $r_{m}$ (taken as BSE-100 index annual returns).
\item Physical risk factor: $PF$.
\item Transition risk factor: $TF$. 
\end{enumerate}
Here, total reported economic losses due to natural calamities for each year are used as a proxy for $PF$. The EM-DAT public worldwide disaster database has been used to assemble information on natural calamities that have occurred in India over the years 1964 to 2022, including the number of occurrences and the associated economic damages \cite{EMDATPublic}. It should be noted, though, that there is no standardized method for estimating economic loss in the EM-DAT database, and different organizations, including disaster management departments in various countries, have developed their own methodologies, to quantify economic losses based, on their particular domains. In its most basic form, the economic harm brought on by a particular catastrophe is estimated using the total damage value at the time of the incident, expressed in USD for the year the natural disaster took place. Apart from this, $TF$ represents our transition risk factor. The $\mathrm{CO}_2$ emissions intensity of the GDP of India is used as the proxy for Transition Risk. The data on this was obtained from the database of the International Energy Agency (IEA), which is an intergovernmental organization that aims to promote clean, secure, and sustainable energy for all \cite{IEA2021}. It was established in 1974 in response to the oil crisis and currently has 30 member countries. The IEA provides energy analysis, forecasts, and policy recommendations to its member countries and the global community. The results of the regression analysis carried out for the green portfolio returns model are presented in Figure \ref{fig 6.1}.
In the Table presented in this figure, the ``const'' coefficient value represents $\alpha$, ``ind'' represents the excess market return ($r_{m}-r_{f}$), ``pf'' represents the physical risk factor, and ``tf'' represents the transition risk factor. Further, the second column shows the corresponding coefficients. It can be observed that the corresponding $P$-values for each of these variables is below $0.05$, which means that there exists a statistically significance relationship of the green portfolio return with each of the variables with which regression has been performed. We also see the the model has an $R$-squared value of $0.925$. This means that $92.5\%$ of the variance of the portfolio can be explained by our linear regression model. This is another indicator that the model performs well. In addition, there are several other statistics shown for the regression model which may be observed, but which have not been analyzed separately.

\section{Scenario analysis and stress testing of the green portfolio}

In this section, we observe the behaviour of the green portfolio in various scenarios. For the scenario analysis, it is to be noted that both transition and physical risks can have significant financial implications for companies, investors and governments. Therefore, in order to quantify the climate risk in each scenario, we use a physical risk factor and a transition risk factor. Accordingly, each scenario will be characterized by its physical risk factor and transition risk factor values. This paper also considers a stress scenario, which will have an alarmingly high climate risk present and hence the physical and transition risk factor values will be extreme.

\subsection{Scenario and stress construction}

Three scenarios are being considered in the scenario analysis part of our study. They sequentially represent low/current levels of climate risk, intermediate climate risk and high climate risk. A final stress scenario is considered with an alarmingly high level of climate risk. The period of the next three financial years is chosen for analysis, that is, FY 23-24, FY 24-25, FY 25-26. The predictor variables in our returns model, as described in the preceding section, are projected for each financial year under each scenario. As per the Economic Survey of India 2023, India aims to reduce the emissions intensity of its GDP by $45\%$ from the 2005 levels by the year 2030 \cite{Econsurvey2023}. Assuming the same trend follows for $\mathrm{CO}_2$ emissions intensity of GDP, we should see an annual decrease of emissions by $0.0085$ $\mathrm{tCO}_2$ per USD 1000 in each year from 2023 in the transition risk factor. We categorize this event as having somewhere between intermediate and high climate transition risk. Therefore, we use the average decrease in the past 10 years, that is $0.006$, as the low transition risk factor value. We then take $0.008$ as the intermediate-risk value, $0.0095$ for high risk and $0.0105$ to model the stressed climate transition risk factor. 

In the reference scenario, a baseline rate is chosen for the market to grow in normal/low climate risk conditions. We take this rate as the average of the past 10 years CAGR, the past 15 years CAGR and the past 20 years' CAGR of the BSE 100 index. The market is assumed to grow at the same average rate for each of the three projected years. Similarly, a median value of the past 10 economic losses is chosen for the baseline value of our physical risk factor and it is assumed that this baseline value will be observed each year. The average decrease in the transition risk factor has been $0.006$ for the past 10 years, and the same number reduces the value for every year in the projection period. In the mild risk scenario, the increased physical risk is seen in the FY 24-25 and FY 25-26 projections, where we assume that the damage equals the $70$-th percentile value of damages in the past 10 years. Similarly, the transition risk factor is now decreased yearly by $0.008$. The market is assumed to be largely unaffected by the mild increase in climate risk. Hence, it follows the same projections as the reference scenario for the market index. Coming to the high-risk scenario, the physical risk factor starts on a mildly risky note with a $70$-th percentile value for FY 23-24. It is then projected to attain near the $85$-th percentile damage in FY 24-25 and has near the $80$-th percentile damage values for FY 25-26. The transition risk factor is now decreased yearly by $0.0095$. The market thereby gets affected by the high climate risk. The increased economic losses in FY 24-25 affect the market and reduce its annual returns for that year. FY 25-26 faces bad losses as well, but the market will show a bounce-back effect, too, from FY-24-25. Therefore, a baseline market return is assumed for FY 25-26. Finally, in the stress scenario, the physical risk factor shows a high, near $75$-th percentile value in FY 23-24. The risk really creeps up in FY 24-25 with an extreme physical risk event, leading to a loss of USD 25 billion USD (near $95$-th loss percentile). Harsh conditions of physical risk are prevalent in FY 25-26 as well, with an above $80$-th percentile economic loss. The transition risk factor now decreases sharply by $0.0105$ yearly. The market is bound to be affected by the extreme physical risk event, and hence it shows a $7.5\%$ decline in FY 24-25. The next year faces bad losses due to physical risk events as well, but the market will show a bounce-back effect, too, from FY 24-25. Therefore, an above-baseline market return is assumed for this year. 

\begin{figure}[h]
\begin{center}
\includegraphics[scale=0.65]{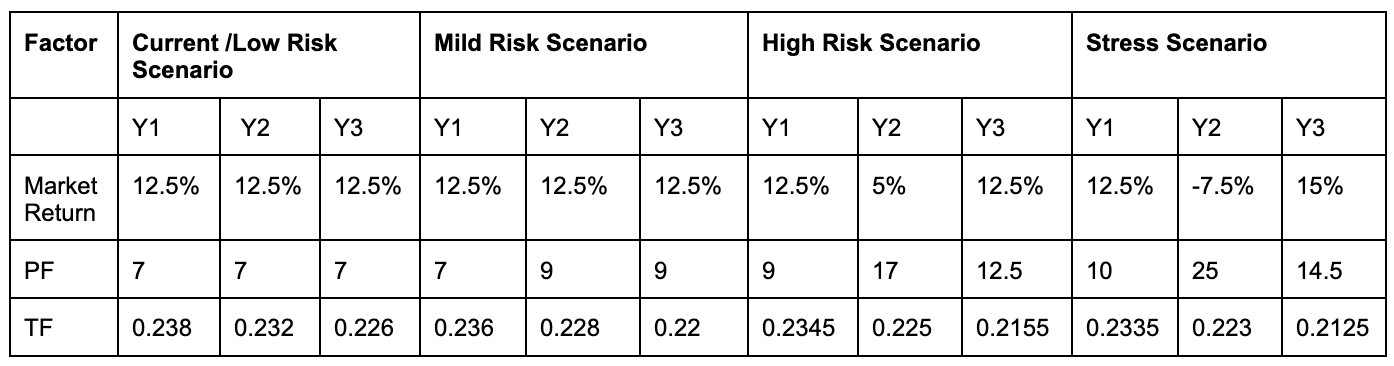}
\caption{Predictor variable projections in each scenario}
\label{fig 7.1 }
\end{center}
\end{figure}

Based on the explained reasoning, presented in Figure \ref{fig 7.1 } are the scenarios and the values our predictor variables are projected to have in each of the scenarios. Here Y1, Y2 and Y3 correspond to FY 23-24, FY 24-25 and FY 25-26, respectively. Further, PF is the physical risk factor and TF is the transition risk factor, with the
units of PF and TF being USD Billion and $\mathrm{tCO}_2$ per USD 1000, respectively.

\subsection{Returns and performance analysis}

\begin{figure}[h]
\begin{center}
\includegraphics[scale=0.6]{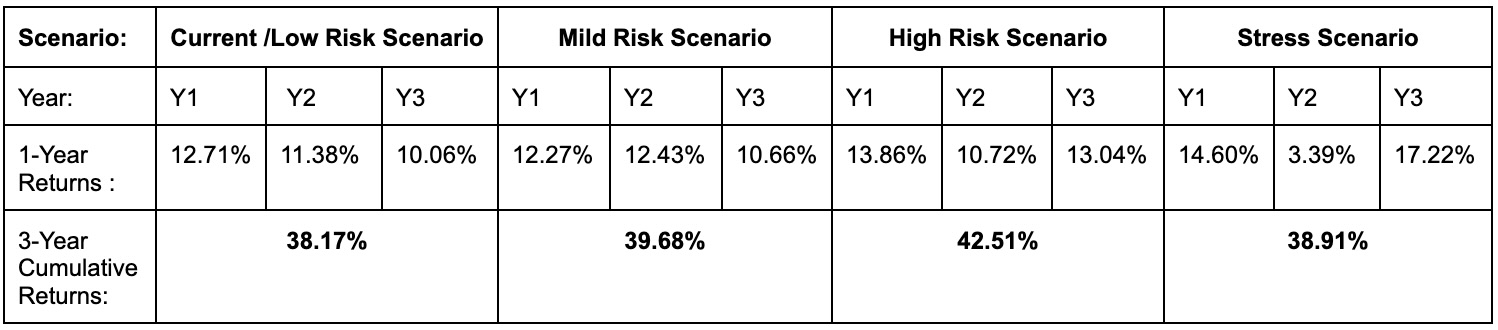}
\caption{Results of scenario analysis}
\label{fig 7.2 }
\end{center}
\end{figure}

\begin{figure}[h]
\begin{center}
\includegraphics[scale=0.6]{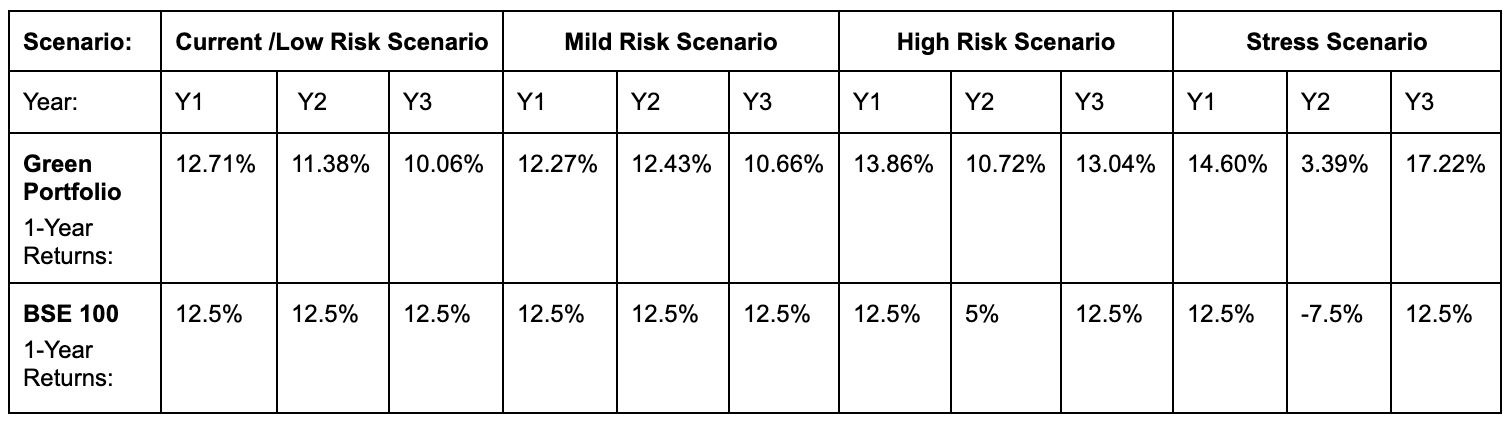}
\caption{Yearly comparison with market index in future scenarios}
\label{fig 8.1}
\end{center}
\end{figure}

\begin{figure}[h]
\begin{center}
\includegraphics[scale=0.65]{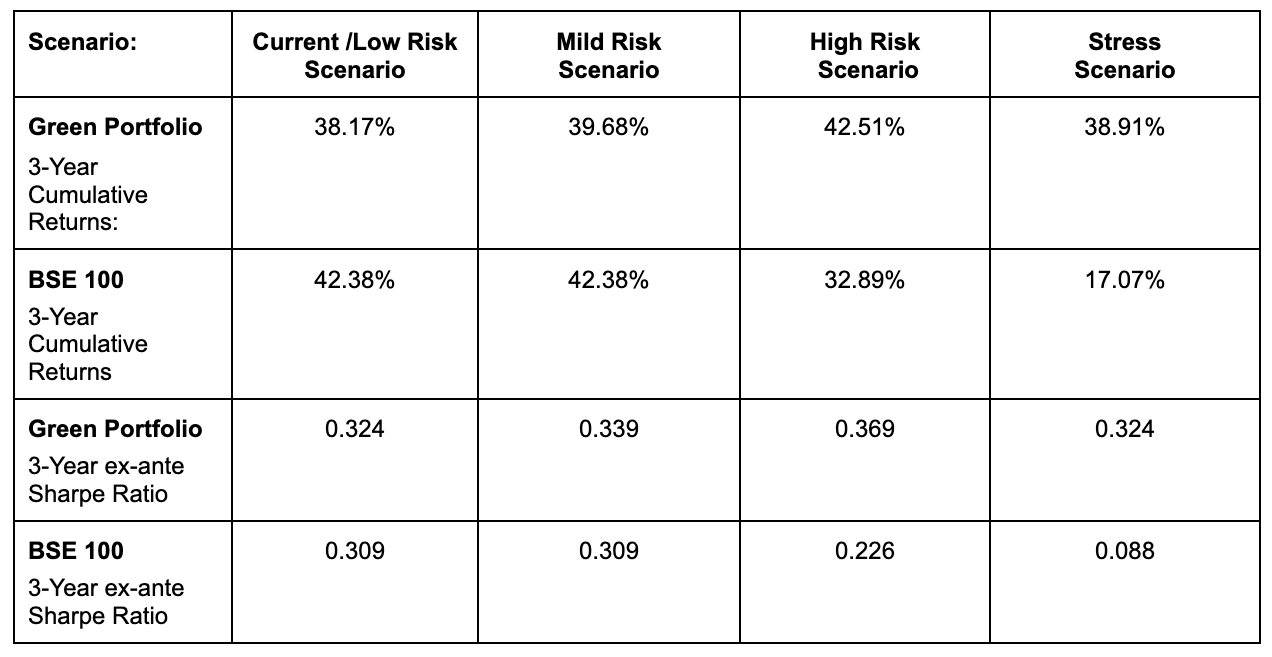}
\caption{3-Y Cumulative Returns and 3-Y ex-ante Sharpe ratios}
\label{fig 8.2}
\end{center}
\end{figure}

Using the predictor variables in each scenario, the portfolio returns for each of the three future years are projected through the returns model used for regression displayed below: 
\[r_{V}=r_{f}+\alpha+\beta_{1}(r_{m}-r_{f})+\beta_{2}PF+\beta_{3}TF.\]
The cumulative 3-year returns using each of the three one-year returns are also calculated and presented in Figure \ref{fig 7.2 }, which is the result of our scenario analysis.
In the Figure \ref{fig 8.1}, we present the annual returns for each of the three years in the projection period in each scenario. We present projections for both the green portfolio and the BSE 100 Index, side by side.
The projected three-year cumulative returns are of greater importance for performance analysis. The analysis also looks at the projected risk-adjusted returns in the form of the ex-ante Sharpe ratios. The ex-ante Sharpe ratio is a measure of risk-adjusted return that is calculated using expected returns and expected volatility rather than historical returns and volatility. We keep the volatility value here to be equal to the historical volatility for simplicity. The results are presented in Figure \ref{fig 8.2} to analyze the same.
In Figure \ref{fig 8.2} we see that the cumulative three-year returns projected to be provided by the green portfolio in low and mild climate risk scenarios are similar to the returns projections for the BSE 100 market index. But, when the climate risk creeps up in the high climate risk scenario, we can see a considerable difference between the projected returns of the green portfolio and the market return. The difference further widens in the stress scenario with alarmingly high climate risk. The ex-ante Sharpe ratios further support the claim. On a risk-adjusted basis, the green portfolio is expected to outperform the BSE-100 index in all scenarios, with the difference rising as the climate risk increases from a low to an alarmingly high level in the stress scenario. This shows the expected resiliency of the green portfolio to climate risk as compared to the current BSE-100 market index.

\section{Conclusions}

We finally summarize the main conclusions and takeaways from the comprehensive analysis presented in the paper:
\begin{enumerate}[label=(\Alph*)]
\item The classical two-dimensional frontier is extended by adding a third dimension of environmental friendly practices (as reflected by the environmental component of the ESG scores) to create a three-dimensional green efficient frontier. The green returns are maximized for determining an optimal green portfolio.
\item Physical and transition risk factors can be created using simple proxy indicators like annual total economic losses due to natural calamities and emissions intensity of GDP. Advanced metrics can further be used to create sophisticated physical and transition risk scores for companies and portfolios.
\item The constructed green equity portfolio gives superior projected returns compared to the market when climate risk exceeds the intermediate levels. The difference in projected returns increases as climate risk increases. Projected risk-adjusted returns are expected to exceed market risk-adjusted returns in all scenarios.
\item Advanced metrics assessing a company's climate risk vulnerability, assessing exposures such as location, resource dependence, and regulatory adaptability, can enhance stress testing and scenario analysis for financial portfolios. This enables better risk mitigation and identification of sustainable investment opportunities.
\end{enumerate}

\section*{Declaration of Interests}

The authors declare that they have no known competing financial interests or personal relationships that could have appeared to influence the work reported in this paper.

\bibliographystyle{elsarticle-num}

\bibliography{BIBLIO}
	
\end{document}